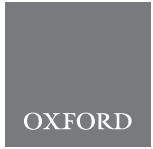

## ORIGINAL ARTICLE

# Neurofeedback Tunes Scale-Free Dynamics in Spontaneous Brain Activity

T. Ros[1], P. Frewen[2], J. Théberge[3], A. Michela[1], R. Kluetsch[4], A. Mueller[5], G. Candrian[5], R. Jetly[6], P. Vuilleumier[1,†], and R. A. Lanius[2,†]

[1]Geneva Neuroscience Center, Department of Neuroscience, University of Geneva, CH-1202 Geneva, Switzerland, [2]Department of Psychiatry, Western University, London N6A 5A5, Ontario, Canada, [3]Department of Medical Imaging, Lawson Health Research Institute, London N6C 2R5, Ontario, Canada, [4]Department of Psychosomatic Medicine and Psychotherapy, Mannheim-Heidelberg University, 68159 Mannheim, Germany, [5]Brain and Trauma Foundation, CH-7000 Chur, Switzerland, and [6]Directorate of Mental Health, Canadian Forces Health Services, Ottawa K1A 0K6, Canada

Address correspondence to T. Ros, Department of Neuroscience, University of Geneva, Geneva, Switzerland. Email: tomasino.ros@gmail.com

[†]These authors contributed equally to this work.

## Abstract

Brain oscillations exhibit long-range temporal correlations (LRTCs), which reflect the regularity of their fluctuations: low values representing more random (decorrelated) while high values more persistent (correlated) dynamics. LRTCs constitute supporting evidence that the brain operates near criticality, a state where neuronal activities are balanced between order and randomness. Here, healthy adults used closed-loop brain training (neurofeedback, NFB) to reduce the amplitude of alpha oscillations, producing a significant increase in spontaneous LRTCs post-training. This effect was reproduced in patients with post-traumatic stress disorder, where abnormally random dynamics were reversed by NFB, correlating with significant improvements in hyperarousal. Notably, regions manifesting abnormally low LRTCs (i.e., excessive randomness) normalized toward healthy population levels, consistent with theoretical predictions about self-organized criticality. Hence, when exposed to appropriate training, spontaneous cortical activity reveals a residual capacity for "self-tuning" its own temporal complexity, despite manifesting the abnormal dynamics seen in individuals with psychiatric disorder. Lastly, we observed an inverse-U relationship between strength of LRTC and oscillation amplitude, suggesting a breakdown of long-range dependence at high/low synchronization extremes, in line with recent computational models. Together, our findings offer a broader mechanistic framework for motivating research and clinical applications of NFB, encompassing disorders with perturbed LRTCs.

## Introduction

The multiple spatiotemporal scales through which brain activity can be studied present a veritable challenge in neuroscientists' quest to link brain and behavior. Although traditional approaches have been mostly restricted to discrete spatial or temporal scales, recent investigations are bearing witness to an emerging interest in scale-free (or "fractal") measures of brain function (He et al. 2010; Kello et al. 2010). As their name implies, such measures are able to capture relationships "across" different levels of brain organization, inherent for example, in the temporal structure of electrocortical patterns (Van de Ville et al. 2010) or the topology of functional connectivity networks (Liu et al. 2014). Here, scale-free relationships are frequently represented by the Hurst scaling exponent (*H*), as a measure of self-similarity within timeseries, which indicates the degree of long-range temporal (auto)correlations (LRTCs) present between shorter and longer timescales (e.g., ranging from seconds to minutes). A larger *H*-value generally reflects the presence of a long-range (yet transient) trend in the data, for example, when values alternate between high and low values, but do so for a prolonged period at a time in each state. Given that *H* integrates a signal's temporal evolution, it may be regarded to index its long-term dependence, or "memory." Thereby, *H* essentially estimates the extent of temporal complexity in a signal, with random white noise and a smooth line taking $H = 0.5$ and $H = 1$ values, respectively. A fascinating implication of scale-free indices is their link with the nascent science of complex systems, which reputedly operate at a "critical" balance between spatiotemporal order and disorder (Chialvo 2010). Intriguingly, spontaneous neural activity has been found to exhibit temporal memory reflected in positive long-range dependence (i.e., $H > 0.5$) (Linkenkaer-Hansen 2001; Nikulin and Brismar 2004), lending support to hypotheses that the brain may display a "self-organized" form of criticality (SOC) (Bak et al. 1987; Hesse and Gross 2014): a homeostatic state that maximizes the dynamic range and memory required for information processing (Hsu and Beggs 2006; Shew and Plenz 2012). This notion is consistent with studies in patients with brain disorders, which report abnormal resting-state *H*-values relative to the healthy population, thus signaling disrupted LRTCs. For example, schizophrenia (Nikulin et al. 2012) and Alzheimer's disease (Montez et al. 2009) patients exhibit attenuated (i.e., more random) LRTCs in the amplitude fluctuations of alpha-band oscillations, while theta-amplitude LRTCs seem to negatively correlate with major depression severity (Linkenkaer-Hansen et al. 2005). Elsewhere, LRTCs have been directly linked to fluctuations in behavioral performance (Palva et al. 2013; Smit et al. 2013), brain maturation (Smit et al. 2011; Iyer et al. 2015), and levels of consciousness (Tagliazucchi et al. 2013; Barttfeld et al. 2015). Notably, both computational modeling (Poil et al. 2012) and subdural recordings (Monto et al. 2007) suggest that network changes in the excitation/inhibition balance may underpin disordered LRTCs.

In light of this interdependence between temporally-complex neural dynamics, behavior, and pathology, a therapeutic means by which to normalize LRTCs might unlock a promising treatment avenue for brain disorders, collectively the leading contributor to global disease burden (Collins et al. 2011). In this work, we report on the exciting possibility of tuning LRTCs of spontaneous brain oscillations following non-invasive, closed-loop neurofeedback (NFB) training (Kamiya 2011). During NFB, a sensory description of real-time brain activity is fed-back to users via a brain–computer interface (BCI), enabling top–down control of network oscillations, including amplitude (Hardt and Kamiya 1978), frequency (Angelakis et al. 2007), and functional connectivity (Brunner et al. 2006). Although NFB is showing early promise for treating brain disorders (Niv 2013; Ros et al. 2014), there is still a poor mechanistic understanding of its influence on brain function. Interestingly, as NFB has been shown to concomitantly alter the amplitude of cortical oscillations and the balance of cortical excitation/inhibition (Ros et al. 2010), we recently hypothesized this might be associated with changes in signature(s) of critical brain dynamics (e.g., LRTCs) (Ros et al. 2014). In this work, we begin by demonstrating that one session of NFB training may reliably enhance alpha-band LRTCs in healthy subjects relative to a sham-control group. We then go on to show that NFB is able to rescue abnormally reduced alpha-band LRTCs in patients with post-traumatic stress disorder (PTSD), together with a significant and correlated improvement in symptoms of hyperarousal.

## Materials and Methods

### Subjects

Experiments were approved by the Research Ethics Board of Western University, Ontario, Canada and all participants were recruited from the university neighborhood. "Experiment 1" involved a group of 40 healthy adult subjects (mean age 33.6, standard deviation (SD) 11.1, 25 female). These participants were carefully screened for the presence of neurological or psychiatric disorders during a structured Structured Clinical Interview for DSM-IV Axis I Disorders (SCID-I) Interview. "Experiment 2" involved 21 adult PTSD patients (mean age 39.9, SD 13.7, 18 female). All met the DSM-IV (American Psychiatric Association and Task Force on DSM-IV 2000) criteria for a primary diagnosis of PTSD related to childhood maltreatment. Axis I diagnoses were assessed by a trained psychologist using the SCID-I (First et al. 1995) and the Clinician-Administered PTSD Scale (CAPS, total cut-off score >50) (Blake et al. 1995). The CAPS indicated that the PTSD group met the necessary criteria for trait hyperarousal (a score of >2/5). Exclusion criteria comprised a lifetime diagnosis of a psychotic disorder, bipolar disorder, substance use disorders, a history of head trauma, serious medical or neurological illness. Eleven participants were currently taking psychotropic medications. The healthy control group consisted of 30 healthy adult subjects (mean age 39.4, SD 8.7, 26 female), and was matched for age/gender from the Human Brain Institute (HBI) normative database (http://www.hbimed.com/). Lastly, an additional set of healthy adults ($n = 32$) were recruited for a resting-state "eyes closed" recording. All participants' electroencephalogram (EEG) recordings were performed using the same amplifier type (Mitsar-201).

### Experimental Timeline

For all NFB, sham-neurofeedback (SHAM), and PTSD participants, the overall protocol consisted of 3 sequential parts that occurred within the same daytime visit: fMRI scan before NFB (~30 min), EEG-NFB (30 min), and fMRI scan after NFB (~30 min). fMRI analyses of Experiments 1 and 2 have been previously reported in Ros et al. (2013) and Kluetsch et al. (2014), respectively, while we additionally recruited another 6 participants to boost the overall sample of Experiment 1. In this article, we focus on the EEG resting-state recordings that directly flanked the start and end of one individual NFB session. Upon arrival to the examination facility, healthy participants were randomized to 1 of 2 experimental groups: EEG-NFB (NFB, $n = 20$) or sham-neurofeedback (SHAM, $n = 20$). For ethical reasons, PTSD patients received only EEG-NFB. Each session consisted of a 3-min resting state (no-feedback), followed by 30 min of continuous feedback (real or sham), and lastly another 3-min resting state (all in eyes open). Immediately before and after the resting state, participants completed Thayer's Activation–Deactivation Checklist questionnaire. Here, the "uncalmness" subscale was used as a self-reported measure of state arousal. No adverse effects were reported after NFB or SHAM feedback.

### EEG Recording and Feedback Training

Scalp voltages of all participants were recorded using a 19 channel electrode cap (Electro-cap International, Inc.) according to the 10-20 international system, with ground electrode at AFz and linked-ear reference. Electrical signals were amplified with the Mitsar 21-channel EEG system (Mitsar-201, CE0537, Mitsar, Ltd) and all impedances were kept under 5 kΩ. EEG was recorded continuously, digitized at a sampling rate of 250 Hz, and stored on hard disk for offline analysis. In parallel, a bridged Pz channel was specifically used for NFB and was connected to a ProComp+ amplifier (Thought Technology) interfacing with EEGer 4.2 neurofeedback software (EEG Spectrum Systems) with right and left earlobes as ground and reference electrodes, respectively. During (feedback-free) resting-state recordings, participants were asked to relax with their eyes open and gaze at a blank wall. SHAM group participants did not receive veridical feedback from their brain activity, but were replayed EEG signal from a previously recorded session of a NFB-successful participant (their whole-scalp EEG activity was nevertheless recorded passively). For online training, the EEG signal was IIR (infinite impulse response) band-pass filtered to extract alpha (8–12 Hz) amplitude with an epoch size of 0.5 s. Here, subjects were rewarded upon reduction of their absolute alpha amplitude, where threshold for reward was set to occur 60% of the time below the initial 3-min baseline average (i.e., 40% negative feedback). Visual feedback was clearly displayed on a 17″ monitor via 1) a dynamic bar graph at the center of the screen whose height was proportional to real-time alpha fluctuations and 2) a "Space Race" game, where a spaceship advanced through space when amplitude was below threshold, and became stationary when above threshold. No explicit instructions were given on how to achieve control over the spaceship, and all participants were told to be guided by the visual feedback process. The 30-min session was divided into ten 3-min periods, with a break of 10 s between each period.

### EEG Analysis

For offline analyses, EEG signals were re-referenced to common-average reference. Low- and high-pass filters were set to 0.5 and 40 Hz, respectively, with a 55–65 Hz notch filter. EEG data were analyzed with the Neurophysiological Biomarker Toolbox (NBT, http://www.nbtwiki.net/) in Matlab (MathWorks Inc.). We used ICA decomposition to first remove stereotypical artifacts using the Infomax algorithm (blinking and lateral eye movement). Statistically defined artifacting was then carried out with the FASTER plug-in (Nolan et al. 2010) removing segments based on extremal deviations of amplitude and variance from the mean. Based on the NFB protocol, our hypotheses focused on 1) alpha amplitude LRTC and 2) mean alpha amplitude. We restricted LRTC analyses to the resting state given that training data were more contaminated and of variable length between subjects when artifacted. Each 3-min EEG recording was firstly decomposed with a discrete wavelet transform (Tumari et al. 2013; Kitzbichler and Bullmore 2015) using the Wavos toolkit (Harang et al. 2012). Here, a Daubechies 4-tap wavelet was used and D5 detail extracted, corresponding to 7.8–15.6 Hz (i.e., alpha-band). Then, the instantaneous alpha amplitude (envelope) across time was determined using the absolute value of the Hilbert transform (see Fig. 1). LRTCs were subsequently estimated via detrended fluctuation analysis, which estimates the scaling of the root-mean-square fluctuation of the integrated and linearly detrended signals, F(t), as a function of time window size, t (see Hardstone et al. 2012 for more details). As shown in Figure 2, H (Hurst) exponents were calculated using a linear regression fit on a log–log graph with a range of 10 equally spaced points between 1 and 30 s, in line with previous work (Linkenkaer-Hansen 2001; Montez et al. 2009). A linear goodness-of-fit was additionally calculated for each DFA estimation using the R-squared statistic (Nikulin and Brismar 2004). See Supplementary Material for mathematical detail on the method. Lastly, absolute alpha amplitude (8–16 Hz) was estimated with a standard FFT approach using Welch's method (Matlab "pwelch" function) and a Hanning windowing function (4 s epoch, 50% overlap). Relative alpha amplitude was calculated as the ratio of the mean alpha amplitude and the broadband amplitude (1.5–40 Hz).

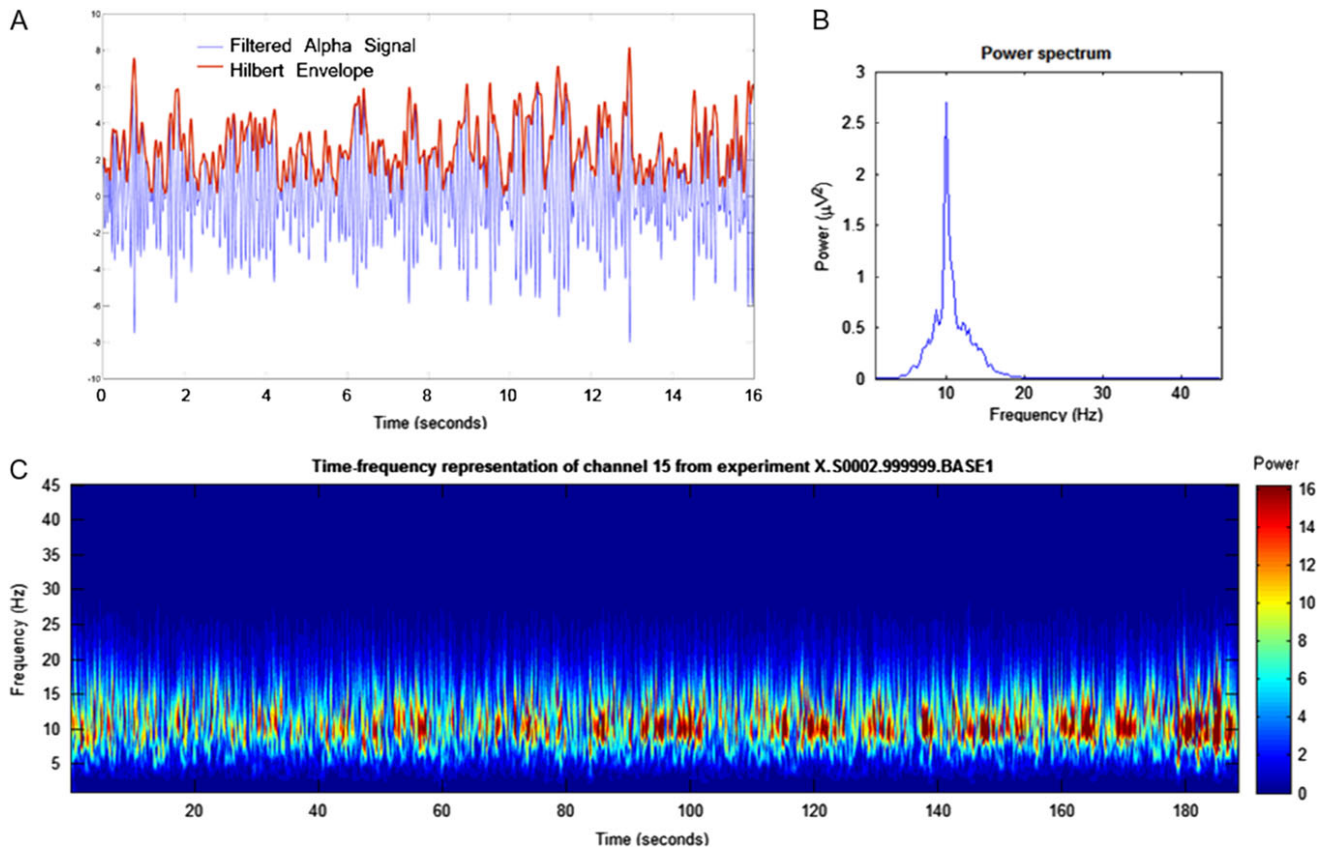

Figure 1. Summary of signal processing steps. (A) Wavelet-filtered alpha oscillation (blue trace) recorded from Subject #2 (Pz channel) during resting state, including its instantaneous amplitude envelope (red trace). (B) Power spectrum of the same alpha oscillation. (C) Instantaneous alpha amplitude (red trace) can also be visualized on a wavelet time-frequency plot where bursts of high/low alpha amplitude (red/blue, respectively) are evident during the whole resting-state recording. LRTCs of alpha amplitude fluctuations are then estimated for the entire timerange of 3 min.

### Statistical Analysis

Multivariate analysis of variance (MANOVA, Pillai's trace statistic) F-tests were used to assess global changes of alpha amplitudes and H exponent. The within-subject factors were "Time" ($T_1$ and $T_2$, i.e., pre and post feedback) and "Channels" (19 channels, see above). The between-subject factor was "Group" (NFB and SHAM). In order to account for multicollinearity among EEG channels, multivariate Hotelling's $T^2$ tests were conducted to examine directionality of the "Time" ($T_1$ and $T_2$) and "Group" (NFB and SHAM, or PTSD and CONTROLS) factors. Secondly, in order to reveal topographical (univariate) effects, 2-tailed Student t-tests were performed channel-wise, within and between subjects (dependent and independent tests, respectively); P-values below 0.05 and 0.01 are indicated on topographic plots. Effect sizes are reported as partial eta squared ($\eta^2$). Following Montez et al. (2009), a correction for multiple comparisons was not necessary, because the number of channels with P-values below 0.05 ranged from 5 to 10 and the likelihood of having this many channels out of 19 by chance was less than 2% (cf., binomial distribution). Moreover, the channels were anatomically clustered in the topographic plots. Grand-average amplitudes and H-exponents across "all channels" are reported with standard error of mean (SEM) for respective experimental conditions and subject groups. Group differences in these grand-average means were computed using a 2-tailed t-test with P < 0.05. Lastly, least-squares correlations were estimated using the Pearson coefficient R. All analyses had a statistical significance threshold of alpha = 0.05, and all post hoc comparisons were Bonferroni corrected.

## Results

### Experiment 1: Investigating Real Versus Sham NFB on LRTCs in Healthy Subjects

We investigated whether NFB regulation of the dominant human EEG rhythm, the alpha oscillation, would reliably alter alpha-band LRTCs in the post-NFB resting state. Here, we provided real-time feedback of alpha fluctuations on a computer visual display, and asked subjects to volitionally attenuate their parietal alpha amplitude for a total of 30 min; "before" and "after" which a 3-min full-scalp EEG was recorded in the eyes open condition. This procedure was randomized to 2 groups: an experimental group (n = 20) receiving veridical (i.e., real time) feedback (NFB), and a control group (n = 20) receiving sham (i.e., pre-recorded EEG) feedback (SHAM).

As shown in Figure 1, and in accordance with prior work (Linkenkaer-Hansen 2001), LRTCs were estimated from the Hilbert envelope (i.e., amplitude) of wavelet-filtered alpha oscillations (8–16 Hz) over a resting-state period of 3 min. For each channel timeseries, the Hurst (H) scaling exponent was estimated using DFA, and computed as the slope of the linear regression line on a log–log plot, with temporal windows ranging from 1 to 30 s (Fig. 2). In accordance with previous work (Nikulin and Brismar 2004), the linear model accounted for >99.5% of the variance explained in the entire data set, while no significant differences of goodness-of-fit were present between experimental groups or conditions ($P < 0.05$), excluding the possibility that group/condition differences in $H$ exponent were driven by departures from the linear model.

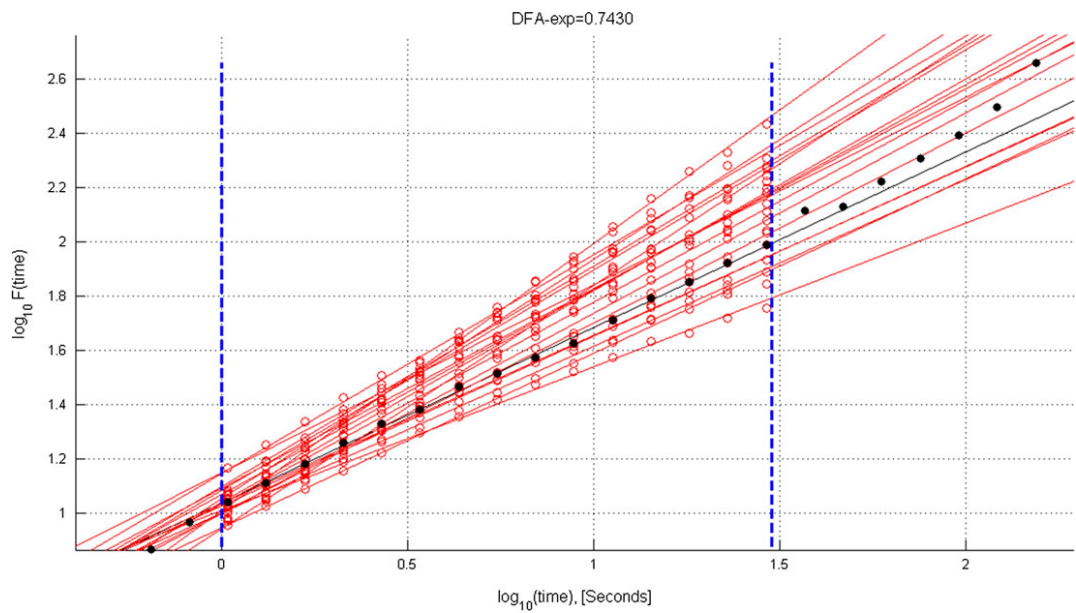

Figure 2. Graphical representation of DFA. The Hurst exponent was defined by the slope of a linear regression line on a log–log plot, fitted within a min–max range of different window sizes (blue vertical lines). Here, the slope of the DFA log–log plot is numerically equivalent to the Hurst (H) exponent for values <1. For illustration, H exponents (red lines) were calculated here for NFB-group subjects at baseline (Pz channel), with a fitting interval of 1–30 s; the black line highlights the fit for subject #11 (H = 0.74).

One-way MANOVAs ("Group") of NFB and SHAM groups showed no significant "baseline" ($T1$) differences in overall $H$ exponent ($F_{19,20} = 1.4$, n.s.), nor alpha amplitude ($F_{19,20} = 1.0$, n.s.). Regarding outcomes of feedback training, a 2-way MANOVA revealed a significant "Group × Time" interaction in $H$ exponent ($F_{1,38} = 5.9$, $P < 0.05$, $\eta^2 = 0.14$), demonstrating a dissociation in pre-to-post feedback resting-state values between NFB and SHAM groups. As seen in Figure 3A,B, testing for directionality using multivariate paired tests ($T_2 - T_1$) indicated a significant increase of $H$-exponent values in the NFB group (Hotelling's $T^2 = 15.8$, $P < 0.05$, $\eta^2 = 0.29$), while no reliable change was present in the SHAM group (Hotelling's $T^2 = 0.5$, n.s., $\eta^2 = 0.013$). On the other hand, as illustrated in Figure 3C,D, there was no "Group × Time" interaction ($F_{1,38} = 0.1$, n.s., $\eta^2 = 0.003$) for mean alpha amplitude, with an absence of significant pre-post differences for NFB (Hotelling's $T^2 = 1.4$, n.s., $\eta^2 = 0.04$) or SHAM (Hotelling's $T^2 = 1.0$, n.s., $\eta^2 = 0.03$) groups. Moreover, Pearson correlation analyses confirmed that pre-to-post shifts in $H$-exponent values were not correlated with intra-individual changes in alpha amplitude, either in NFB ($R = 0.15$, n.s.) or in SHAM groups ($R = 0.09$, n.s.), further demonstrating their independence.

### Experiment 2: NFB Normalization of LRTCs in PTSD

As PTSD is associated with cognitive-affective dysregulation and anomalous EEG signatures (Jokić-Begić and Begić 2003), we reasoned that it may also manifest perturbed LRTCs based on recent evidence in psychological distress (Churchill et al. 2015). We therefore sought to test and replicate the neuromodulatory effects of NFB in 21 adult patients with PTSD. Remarkably, and as shown in Figure 4A,B, paired tests confirmed the hypothesized boost of alpha $H$-exponent values post-NFB (Hotelling's $T^2 = 9.2$, $P < 0.05$, $\eta^2 = 0.19$), which was maximally expressed along midline regions. Mean alpha amplitude experienced a parallel, albeit topographically non-overlapping enhancement ($T^2 = 9.6$, $P < 0.05$, $\eta^2 = 0.19$), as evidenced by Figure 4C,D. Here, global $H$-exponent changes correlated positively with those of global alpha amplitude ($R = 0.72$, $P < 0.05$). In order to clarify whether these values moved toward or away from those of the normal population, we sampled an additional group of 30 healthy adults, matched for age and gender.

As further illustrated in Figure 4, $H$-exponent values for PTSD subjects were found to be significantly lower than CONTROLS at "baseline" (Hotelling's $T^2 = 65.3$, $P < 0.05$, $\eta^2 = 0.57$), indicating relatively more "random" alpha burst dynamics. This novel signature in PTSD patients was accompanied by a deficit in mean alpha amplitude (Hotelling's $T^2 = 112.6$ $P < 0.05$, $\eta^2 = 0.69$), consistent with previous work (Jokić-Begić and Begić 2003). Critically, NFB was found to restore both parameters toward control-group values, thereby reversing previous significant differences ($H$ exponent: Hotelling's $T^2 = 40.5$, n.s., $\eta^2 = 0.45$; Mean alpha amplitude: Hotelling's $T^2 = 56.6$, n.s., $\eta^2 = 0.53$). Thus post-NFB, both dynamical (i.e., LRTCs) and static (i.e., "mean" amplitude) measures of alpha oscillations exhibited a "rebound" from abnormal lows, whereby anomalous baseline differences were restored toward healthy population levels.

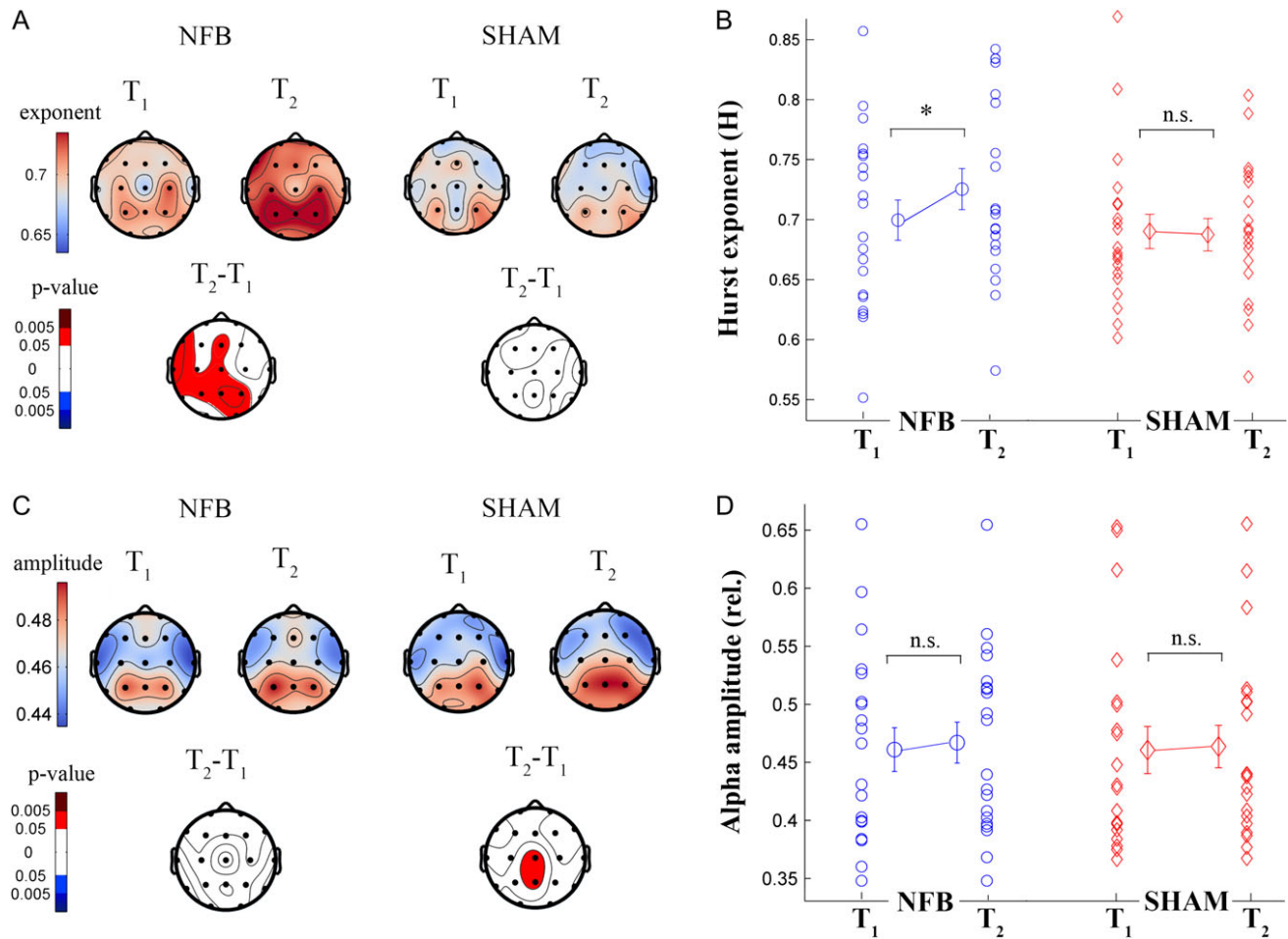

**Figure 3.** Pre ($T_1$) and post ($T_2$) resting-state changes in NFB and sham (SHAM) groups. (A) Topography of Hurst exponents in the alpha-band reflecting LRTCs; (B) mean alpha Hurst exponents; (C) mean alpha Hurst exponents; and (D) mean alpha amplitudes of individual NFB subjects (blue circles) and SHAM subjects (red diamonds). Grand averages indicate all subjects, error bars denote ±SEM. *Significant difference at $P < 0.05$, n.s., not significant.

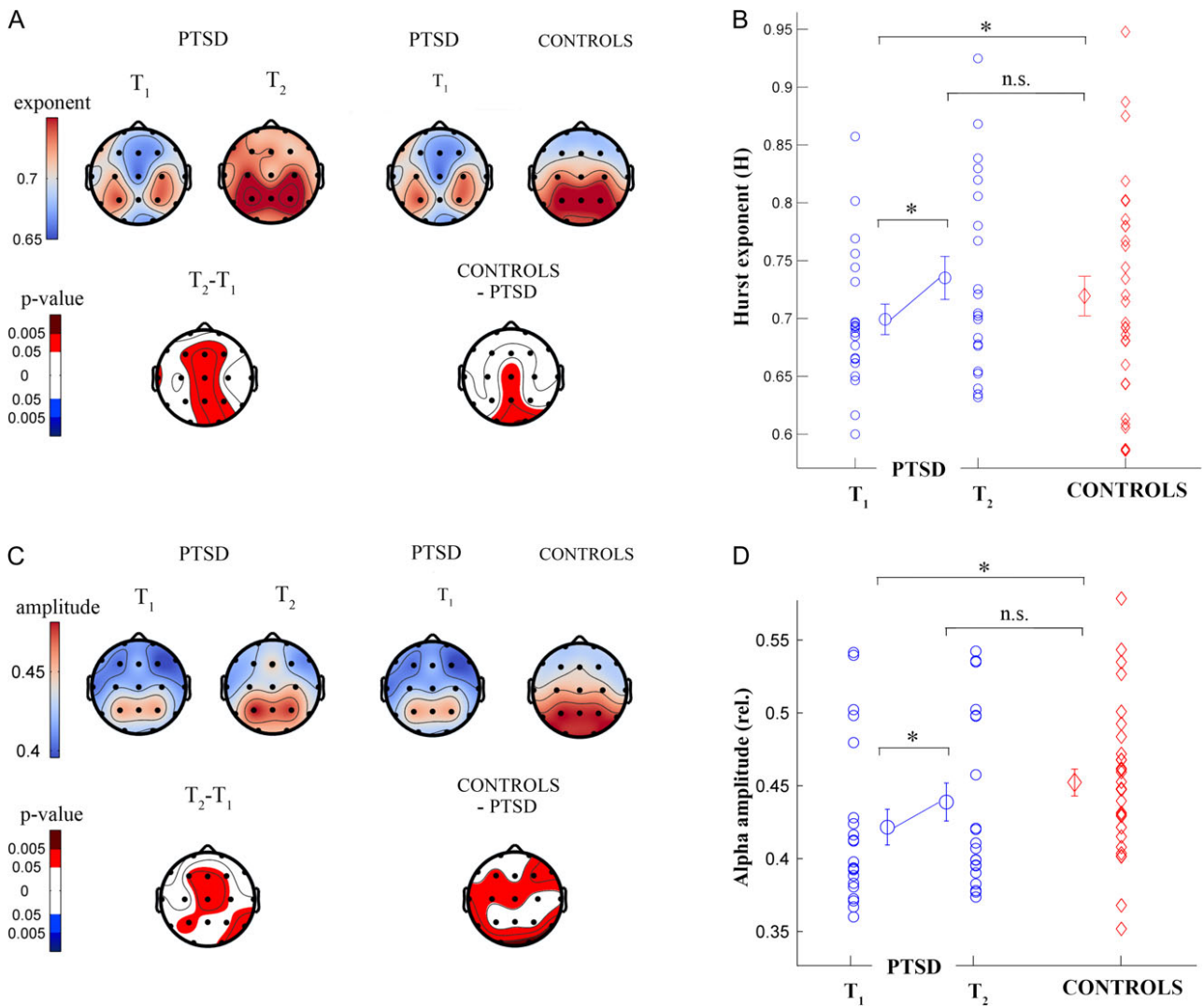

Figure 4. Pre ($T_1$) and post ($T_2$) resting-state changes in PTSD, in comparison with a healthy control group (CONTROLS). (A) Topography of Hurst exponents in the alpha-band reflecting LRTCs; (B) mean alpha Hurst exponents; (C) Topography of alpha amplitudes; and (D) mean alpha amplitudes of individual PTSD patients (blue circles) and CONTROL subjects (red diamonds). Grand averages indicate all subjects, error bars denote ±SEM. *Significant difference at $P < 0.05$; n.s., not significant.

### Changes in Hyperarousal in PTSD Patients

The Clinician-Administered PTSD Scale (CAPS) indicated that the PTSD group met the necessary criteria for trait hyperarousal. A comparison of state arousal (Thayer Activation Checklist) pre-to-post NFB revealed a significant decrease in arousal in the PTSD group ($t = -2.72$; $P < 0.05$) after NFB training. Crucially, as depicted in Figure 5, individual decreases in arousal score were significantly correlated with increases in Hurst exponent ($R = -0.45$, $P < 0.05$) at the feedback channel Pz. A similar (marginally-significant) trend was identified for mean alpha amplitude ($R = -0.39$, $P = 0.08$).

### Relationship Between LRTCs and Mean Alpha Amplitude

We reasoned that a large span of individual data could prove crucial for uncovering a potential non-linear relationship (Botcharova et al. 2014; Teixeira and Shanahan 2015) between LRTCs and oscillation amplitude, as the latter is known to covary with different levels of neural synchronization (Musall et al. 2014). Hence, we collected and analyzed an additional sample comprising eyes-closed recordings (32 healthy subjects, "before" and "after" NFB). Combining data from all experiments ($n = 123$ subjects) on a single scatter plot, we conducted a regression analysis to estimate the best curve-fit. As illustrated in Figure 6, a quadratic fit provided the best model ($R^2 = 0.21$, $P < 0.05$), explaining around 20% of the total variance. This quadratic relationship held for "absolute" alpha amplitude ($R^2 = 0.24$), indicating a superior fit ($Z = 2.03$, $P < 0.05$) versus a linear relationship ($R^2 = 0.11$), which would result from a simple increase in signal-to-noise arising with higher alpha amplitudes (Linkenkaer-Hansen 2001). The statistical presence of an inverted-U relationship is striking, revealing a middle zone of maximal LRTCs, and attenuated LRTCs at higher and lower alpha extremes.

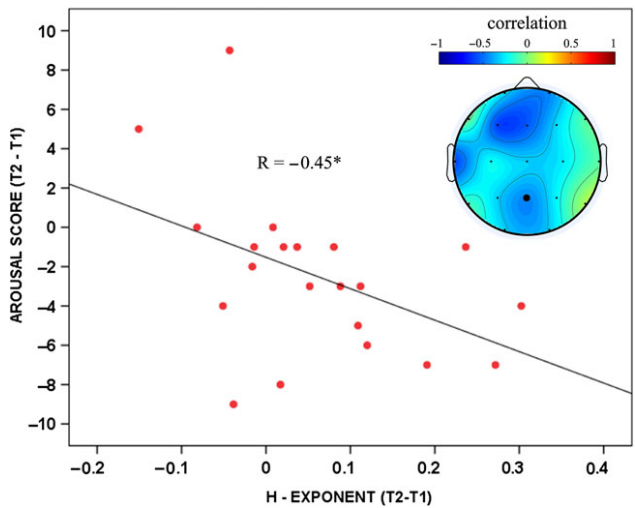

**Figure 5.** Changes in state arousal versus LRTCs in PTSD patients, pre-to-post NFB. Scatter plot of individual change in state arousal versus alpha Hurst exponent change at parietal feedback channel Pz. Inset: whole-scalp distribution of correlations (large dot denotes feedback channel Pz). *Significant Pearson correlation at $P < 0.05$.

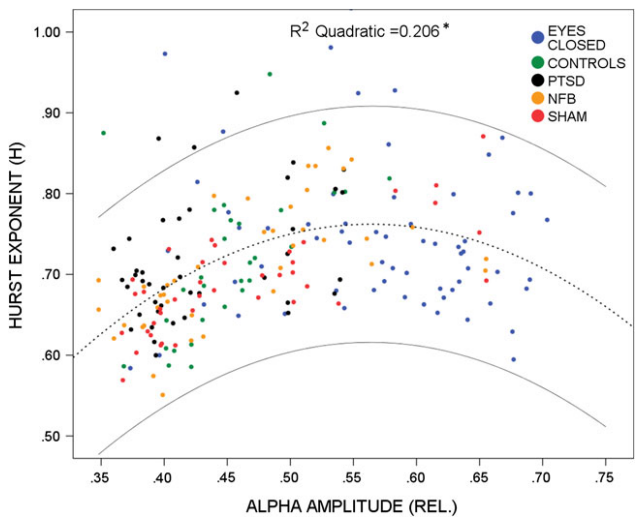

**Figure 6.** Relationship between individual values of global Hurst exponent and global alpha amplitude (i.e., mean of all channels). Colors represent different subject groups: real-NFB, sham-feedback (SHAM), PTSD, PTSD healthy control subjects (CONTROLS), extra eyes closed subjects (EYES CLOSED). Pre ($T_1$) and post ($T_2$) NFB values are plotted for all subjects (except $T_1$ for CONTROLS only). Dotted line denotes quadratic fit, with upper and lower lines the 95% confidence intervals. *Significant at $P < 0.05$.

### Alpha Training Dynamics During NFB

To confirm that closed-loop training (NFB) resulted in differential changes of the controlled parameter (absolute alpha amplitude), we report the 30-min training session dynamics for the 3 experimental groups: NFB, SHAM, and PTSD. As depicted in Figure 7, and consistent with the NFB protocol, NFB and PTSD groups exhibited a more sustained reduction of alpha amplitudes compared with the SHAM group. For the feedback channel Pz (Fig. 7A,) the alpha amplitude time-course significantly differed between NFB versus SHAM groups (Group × Time interaction: $F_{10, 380} = 3.4$, $P < 0.05$). Similarly, a significant difference was observed between PTSD versus SHAM groups ($F_{10, 390} = 2.7$, $P < 0.05$). Analogous training dynamics were found on the global level, averaging overall channels (Fig. 7B).

## Discussion

Overall, our findings demonstrate that sustained control of brain activity, via NFB, may induce changes in the scale-free dynamics of spontaneous brain oscillations, evidenced by increased scaling exponent(s) of EEG LRTCs. Specifically, a significant post-NFB increase in resting-state alpha-band LRTCs was observed in a group of healthy adults, as compared with a sham-control group that received false feedback (Fig. 3). NFB thereby appears to have induced an adaptive "self-tuning" (Stepp et al. 2015) of spontaneous neuronal dynamics, consisting of a tendency for alpha oscillations to remain alternatively high and low in amplitude for a longer duration of time (Hardstone et al. 2012). Fascinatingly, this result was reproduced in patients with PTSD, where abnormal alpha rhythm dynamics (LRTCs) normalized toward values seen in the healthy population (Fig. 4). Here, the inter-individual degree of LRTC re-organization correlated with reductions in self-reported hyperarousal, which significantly decreased at the group level after NFB (Fig. 5).

### Potential Mechanism of LRTC Modulation and Its Relationship to Brain Function

A question that naturally arises is how scale-free oscillations are linked to brain function and behavior. Recent studies report a temporally direct correlation between LRTC exponents of cortical oscillations and those of individual performance errors (e.g., sensory detection, Palva et al. 2013 and time estimation, Smit et al. 2013). Cortical LRTCs are known to increase during development (Smit et al. 2011), and are more attenuated ("random") in a number of brain disorders, including depression (Linkenkaer-Hansen et al. 2005), schizophrenia (Nikulin et al. 2012), Alzheimer's (Montez et al. 2009), and autism (Lai et al. 2010). Patients with depression also express reduced LRTCs in their sleep (Leistedt et al. 2007b), which are interestingly restored during states of symptom remission (Leistedt et al. 2007a). Hence, self-report data from our PTSD sample (Fig. 5) corroborate observations that normalization of LRTCs appears to track state-related improvements of key psychiatric symptoms. A more provocative implication is that such neuromodulation is able to occur at remarkably rapid timescales (i.e., after only 30 min of training), and do so endogenously, without the need of external agents (e.g., pharmaceutics and electrostimulation). This points to the existence of residual and/or homeostatic plasticity mechanisms in the brain (Marder and Goaillard 2006; Stepp et al. 2015) that may be capable of reversing long-term pathological brain activity. Our results thus pave the way for future work investigating whether these effects can maintain themselves following multiple sessions of NFB training, paralleling evidence of long-term changes in EEG amplitude (Becerra et al. 2006; Gevensleben et al. 2009).

LRTCs may also be over-pronounced, as has been reported in epilepsy (Parish et al. 2004; Monto et al. 2007). Here, down-regulating the excitation to inhibition (E/I) balance via GABAergic agonists seems to dampen the highly correlated oscillations of epileptogenic cortex (Monto et al. 2007). This, together with evidence from computational modeling (Poil et al. 2012), tentatively links LRTC changes to fluctuations in E/I balance. Interestingly, we have previously shown that reducing alpha amplitude via NFB may lead to a plastic increase in cortical excitability, as well as a decrease in intracortical inhibition (Ros et al. 2010). It is plausible that a similar neurophysiological mechanism may be at work here, whereby more sustained (i.e., correlated) oscillations could be subserved by a stronger cortical excitatory drive following NFB (Poil et al. 2012). Yet, the contrary facet remains unresolved: could NFB be equally used to reduce pathologically elevated LRTCs? Here, it is noteworthy that promising applications of NFB in epilepsy already exist (Sterman and Egner 2006; Strehl et al. 2014).

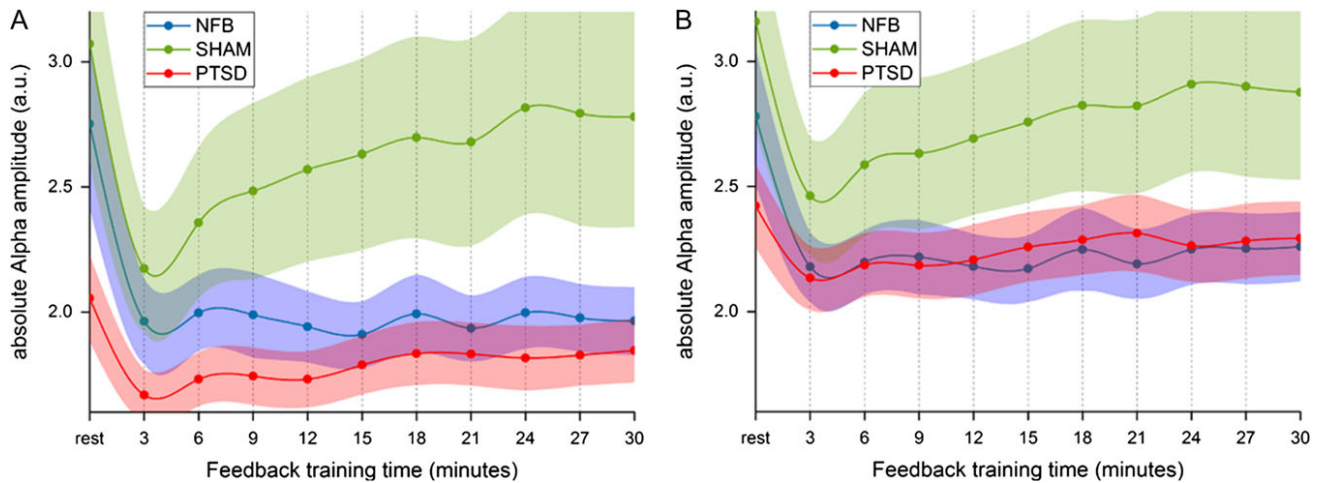

Figure 7. Temporal evolution of alpha amplitude "during" feedback training, for real-NFB healthy subjects (NFB), sham-feedback healthy subjects (SHAM), and PTSD patients. Rest represents the initial 3-min resting-state recording (i.e., $T_1$). The subsequent feedback training was subdivided into 10 periods (3–30 min). (A) absolute alpha amplitude at the feedback site (channel Pz); (B) absolute alpha amplitude globally (i.e., mean of all channels).

### Long-Range Dependence Versus Oscillation Amplitude: A Hidden Link?

The observation of an apparent inverted-U relationship between LRTC magnitude and mean alpha amplitude (Fig. 6) is reassuring as it firstly confirms that LRTC changes cannot be ascribed to a trivial increase in signal-to-noise ratio, as may be expected with ever-greater levels of oscillation amplitude. Secondly, depending on the location of sampled data on the curve, the inverted-U shape provides a parsimonious explanation for the reported differences in association between LRTC and oscillation amplitude: that is, positive (left side) (Poil et al. 2011), null (middle) (Linkenkaer-Hansen 2001), or negative (right side) (Linkenkaer-Hansen 2001). This finding suggests there may be a physiologically limited range of oscillation amplitudes associated with maximal LRTCs. Taking into account several reports indicating that alpha amplitude positively covaries with the synchrony of local neuronal populations (Bollimunta et al. 2011; Haegens et al. 2011; Musall et al. 2014), this suggests that both hypo- and hyper-synchronization appear to be associated with a gradual breakdown of long-range dependence, consistent with theoretical predictions about criticality (Poil et al. 2012; Botcharova et al. 2014; Tomen et al. 2014; Teixeira and Shanahan 2015).

Interestingly, we have found the result to align nicely with a theoretical framework on integrated information in the brain, which suggests a bell-shaped relationship between complexity and neuronal synchronization (Tononi et al. 1998). According to this perspective, complexity as a measure of integrated information (in contrast to plain entropy) is expected to be maximal at intermediate states between high synchronization (order) and low synchronization (disorder). Our data seem to be consistent with this framework (Tononi et al. 1998) when one considers that (alpha) oscillation amplitude frequently tracks the degree of synchronization of intracortical neuronal populations (Bollimunta et al. 2011; Haegens et al. 2011; Musall et al. 2014). Through this lens, LRTCs might be fittingly understood as a multiscale measure of a system's complexity; whereby, once neural oscillations shift to either excessively desynchronized (more disorderly) or synchronized (more orderly) levels there is a parallel shift of LRTCs toward randomness, reflecting a decrease in variability (shallower slope of critical exponent, i.e., dynamic range) and long-range dependence (i.e., memory), in accordance with computational models (Poil et al. 2012). Not coincidentally, critical phenomena such as long-range correlations are found near continuous order-to-disorder transitions (Chialvo 2010; Botcharova et al. 2014; Hesse and Gross 2014; Shanahan and Teixeira 2015), where dynamic range and memory are maximized, both features favorable for information processing (Shew and Plenz 2012). This could be a reason why, in conjunction with deviant LRTCs (Stam et al. 2005; Poil et al. 2011), signatures of hyper- and hypo-synchronization seem to regularly feature in brain disorders (Coburn et al. 2006; Uhlhaas and Singer 2006). It may thus be conceivable that abnormal oscillation amplitudes and LRTCs share a common pathophysiological mechanism (Poil et al. 2008, 2011; Zhigalov et al. 2015). As revealed by our data, PTSD demonstrates both significantly reduced alpha amplitude and LRTCs, which are normalized in tandem post-NFB. Consequently, by directly up or down- regulating oscillation amplitude, in line with conventional NFB approaches (Lubar 1997; Heinrich et al. 2007), one might concurrently trigger an adaptive "self-tuning" (Stepp et al. 2015) of more complex EEG dynamics (i.e., LRTCs), which are scale-free and intractable as a real-time parameter for feedback.

### Could Oscillatory Amplitude and Scale-Free Dynamics be Regulated Homeostatically?

Although our observations demonstrate modulation of LRTCs directly following NFB training, they may be placed in a wider context of studies reporting self-tuning of brain oscillations across the circadian cycle. Intra-individually, EEG oscillation amplitude appears to fluctuate between high (pre-sleep) or low (post-sleep) synchronization levels (Meisel et al. 2013; Plante et al. 2013), a mechanism attributed to synaptic homeostasis. For example, Meisel and colleagues found theta amplitude to consistently increase with sleep deprivation, only to be returned to normal levels after sleep (Meisel et al. 2013). Crucially, this was paralleled by a decrease in synchronization variability during sleep deprivation, which was rescued to baseline levels post-sleep (Meisel et al. 2013). This may be interpreted as initial evidence for homeostatic regulation (i.e., self-tuning) of critical brain dynamics, a dynamic mechanism that appears to be compromised in psychiatric disorder (Plante et al. 2013). Hence, our results complement emerging work by revealing a self-tuning of spontaneous brain oscillations plus LRTCs on much briefer timescales (<1h) and in the absence of sleep, supporting findings that neuronal homeostasis may also occur during waking states (Hengen et al. 2016). This notion is reinforced by recent observations by Zhigalov et al. (2016) showing that closed-loop sensory stimulation—using visual flashes presented at peaks of alpha power—differentially modulates LRTCs during successive sessions of entrainment. This involuntary (i.e., bottom–up) method (Mulholland and Runnals 1962) contrasts interestingly with the voluntary (i.e., top–down) form of NFB based on self-regulation (Kamiya 2011).

### Potential Benefit of NFB in a Range of Brain Disorders

We have shown here the ability of NFB to be used as a neuromodulatory tool in healthy participants as well those with a psychiatric disorder, and have recently put forward a conceptual framework for how top–down training of brain oscillations might help normalize cortical dynamics in a range of psychiatric and neurological conditions (Ros et al. 2014). Although the current study was limited to exploring effects produced by a single training session, earlier investigations have reported a long-term impact of repeated NFB sessions on cortical oscillations (Ros et al. 2014). Given that numerous NFB studies report normalization of targeted EEG amplitudes following long-term treatment, such as attentional-deficit hyperactivity disorder (Arns et al. 2009; Gevensleben et al. 2009), tinnitus (Hartmann et al. 2013), and learning disability (Becerra et al. 2006), we speculate whether this may not have also restored LRTCs in these disorders (see Zhigalov et al. 2015 for their typical distribution across cortical regions). Moreover, normalization of LRTCs could potentially underpin clinical improvements observed with other forms of therapy, be they endogenous, for example, meditation (Lomas et al. 2015) or exogenous (e.g., transcranial alternating current stimulation (Vossen et al. 2015), seeing that many of these treatments have been found to modulate EEG amplitude. For example, transcranial magnetic stimulation (rTMS) in schizophrenia patients has been reported to decrease negative symptoms in proportion to post-treatment increases in alpha amplitude (Jin et al. 2006), while recovery from severe brain injury was found to be associated with reductions of theta amplitude after pharmacotherapy (Williams et al. 2013). Should a unifying mechanism between LRTCs and abnormal oscillation amplitudes be corroborated in future work, it would reinforce the scientific rationale for deploying M/EEG-based NFB in a potentially much wider spectrum of clinical disorders, where it is already demonstrating promise-including schizophrenia (Surmeli et al. 2012), major depression (Escolano et al. 2014), obsessive–compulsive disorder (Kopřivová et al. 2013), insomnia (Schabus et al. 2014), autism (Friedrich et al. 2015), and stroke (Várkuti et al. 2013; Young et al. 2014).

### Conclusion

In summary, our findings establish novel empirical and mechanistic evidence for an intrinsic self-tuning of critical fluctuations following closed-loop training, and directly indicate that NFB could find valuable therapeutic applications in disorders with perturbed LRTCs, such as schizophrenia, major depression, and epilepsy (Linkenkaer-Hansen et al. 2005; Monto et al. 2007; Nikulin et al. 2012).

### Supplementary Material

Supplementary material can be found at: http://www.cercor.oxfordjournals.org/.

### Funding

This work was supported by grants from the EU Marie-Curie CoFund BRIDGE program (grant No. 267171), Lawson Health Research Institute, and Canadian Institute for Military & Veteran Health Research.

### Notes

We thank The Foundation for Neurofeedback and Applied Neuroscience for generously providing the Neurofeedback hardware and software. We are also grateful to Claire Braboszcz for helpful comments, as well as to Suzy Southwell, Stephanie Nevill, Melody Chow, and Nancy Mazza for their assistance with data collection. *Conflict of Interest*: None declared.